\documentclass[aps,prl,twocolumn,floats,nofootinbib]{revtex4}
\usepackage{epsfig}

\begin{document}
\title{Time resolved fission  in metal clusters}
\author{P.~M.~Dinh$^1$, P.-G.~Reinhard$^2$, E.~Suraud$^1,^2$}
\affiliation {$^1$Laboratoire de Physique Th\'eorique,
        Universit{\'e} Paul Sabatier,
        118 Route de Narbonne, F-31062 Toulouse, Cedex 04,
        France}
\affiliation {$^2$Institut f\"ur Theoretische Physik, Universit\"at Erlangen
              Staudtstr.\ 7, D-91058 Erlangen, Germany}

\begin{abstract}
We explore from a theoretical point of view  
pump and probe (P\&P) analysis for fission of metal cluster
where probe pulses are generalized to allow for scanning various 
frequencies.  
We show 
that it is possible to measure the time the system needs to 
develop to scission. This is achieved
by a proper choice of both delay and frequency of the probe pulse.
A more detailed analysis even allows 
to access the various intermediate stages of the fission process. 
\end{abstract}

\maketitle

Pump and probe (P\&P) experiments have become a widely used, versatile
and powerful tool for time resolved studies of dynamical reactions in
all sorts of electronic physics from molecules \cite{Zew94,Gar95} to
bulk \cite{Hoe97}.  Experimental and theoretical investigations of
P\&P analysis for clusters as intermediate size systems have started
more recently and are mostly still on an exploratory level
\cite{Har98b,Lei99,Hei00}, see also \cite{Rei03a} for an
overview. Clusters are here a great challenge
because they are much more complex than molecules and there is thus a
much larger variety of possible scenarios. This calls for theoretical
explorations and there exist already several of them, see
\cite{Rei03a}. In the spirit of the studies in \cite{And02,And03a}, we
propose to 
exploit the Mie plasmon resonance as a handle to track the time
evolution of the cluster shape. A most dramatic evolution of shapes is
experienced in cluster fission \cite{Bre94b,Yan02}, see also \cite{Nae97}
for a review. The
various shapes along the fission path are related to characteristic
resonance spectra \cite{Rei97b} which provides a unique opportunity
for P\&P analysis of the associated time scales. As we shall see
below a standard P\&P scenario, augmented by a frequency scan,
indeed allows to access a fully time resolved analysis of fission
dynamics.
This is an interesting task as such and it carries farther reaching
information.  The time the system needs to attain scission (namely to
actually split into two pieces) provides an indirect but crucial clue
to the viscosity of clusters \cite{Fro97a}.  Fission is also a well
known property in another class of finite fermion systems, namely
atomic nuclei. It is actually by studying fission dynamics in atomic
nuclei that the first experimental estimate 
of nuclear viscosity could be attained
\cite{Dur00}.

The dipole plasmon plays a key role in the dynamics of metal clusters,
both in the linear and non-linear domains \cite{Kre93,Rei03a}. It will again
provide here the handle for the P\&P analysis. The theoretical
modeling needs to account properly for both, electronic and ionic
degrees of freedom. This is achieved by the Time Dependent Local
Density Approximation (TDLDA) for electrons, coupled to Molecular
Dynamics (MD) for ions, the coupling being achieved by a local
pseudo-potential \cite{Kue99}.  Absorbing boundary conditions are used
throughout, which allows a proper description of ionization.  We use
a cylindrically averaged description of the electrons to
simplify the extensive calculations \cite{Cal00}. Details on the TDLDA-MD
approach can be found in \cite{Cal00,Rei03a}.

The laser field is described in the dipole
approximation as a time-dependent external potential
$V_{\rm las} =
{\bf E}_0\!\cdot\!\hat{\bf d} f_{\rm las}(t) \cos(\omega t)$, 
with the  dipole operator $\hat{\bf d}$, the laser polarization and
amplitude ${\bf E}_0$, and $f_{\rm las}(t)$ the temporal profile
of the laser pulse, here chosen as a sin$^2$ profile. Pulses are short
(FWHM typically less than 50 fs) as required for a P\&P scenario.
Laser intensities are kept moderate and tuned to provide the proper
pump excitation and the proper probe analysis in terms of
ionization. Pump as well as probe pulses have the same polarization
but different frequencies and FWHM (see below).

Both electronic and ionic observables are recorded as a function of
time. The two key observables for the P\&P scenario are ionization
and the electronic dipole moment with respect to the ionic center of
mass, which provides the plasmon response by Fourier transformation
into the frequency domain, \cite{Cal00,Yab96}. The ionization is
computed as a function of time, by recording the number of electrons
lost through the absorbing boundaries ($N_{\rm esc}$). Ions are
treated as classical particles by standard MD propagation of positions
and momenta. The P\&P analysis will thus provide a way, with optical
techniques, to access this ionic dynamics experimentally.

\begin{figure}
\epsfig{figure=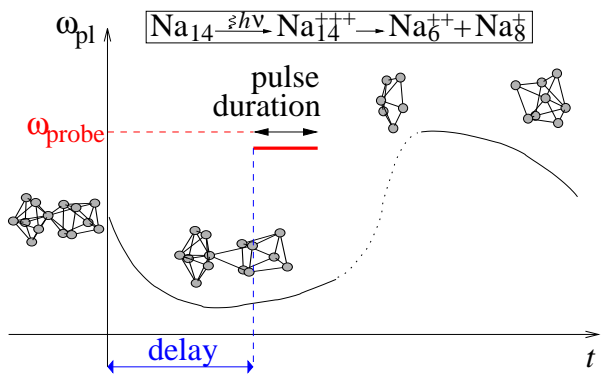,width=8cm}
\caption{\label{fig:scheme}
Schematic drawing of the evolution of the plasmon response of a metal cluster 
along its fission path. Inserts indicate typical ionic structures.
The P\&P analysis has to be considered here in 
terms of both delay and probe frequency as sketched on the figure.
For details, see text.
}
\end{figure}

We consider the cluster Na$_{14}$ initially excited by
a pump laser to 
a 3$^+$ charge state. The laser is linearly polarized along the
symmetry axis of the cluster. The ${\rm Na_{14}}^{+++}$ is unstable
with respect to fission \cite{Nae97} and thus starts to elongate until
it breaks up.  With the pump laser pulse used here (intensity:
$5\times 10^{10}$ W/cm$^2$, frequency: 2.3 eV, FWHM: 36 fs), the
average fission path leads to $\rm {Na_{14}}^{+++} \longrightarrow
{Na_8}^+ + {Na_6}^{++}$. A word of caution is in place here: TDLDA-MD
provides an average description and does not give access to all the
possible fission channels.  Still, even with that limitation TDLDA-MD
suffices for a first exploration of P\&P analysis of cluster
fission. The fission scenario which we have in mind allows a selective
identification of either the fissioning system or the fission
fragments in terms of optical response. It is indeed typical that the
plasmon frequencies of the fragments lie safely 
above the plasmon frequency of the elongated fissioning cluster.  
The key mechanism which makes the analysis
possible is the fact that the ionization yield is strongly correlated
to the plasmon resonance: The closer the frequency of the probe laser
to the plasmon frequency, the larger the ionization and {\it vice 
  versa}~\cite{Ull98a}. The 
plasmon response, in turn, is usually concentrated in a well defined
frequency domain and provides a direct fingerprint of the underlying
ionic structure. Scanning the probe ionization thus provides a strong
clue to the ionic structure and its time evolution.  The position (and
structure) of the plasmon resonance changes in a characteristic manner
along a fission path \cite{Rei97b}: the elongation leads to a
progressive red shift of the plasmon resonance mode along the fission
axis. However, as soon as the system has fissioned, the now more
compact fission products both lead to a significantly blue shifted
response.  This is sketched in figure~\ref{fig:scheme}.  Because of the
expected strong changes of the resonance landscape, the probe pulses
have to hit with the proper frequency as well as the proper delay
time. We will thus consider a generalization of the standard P\&P
scenario to scanning probe frequencies. 

\begin{figure}
\epsfig{figure=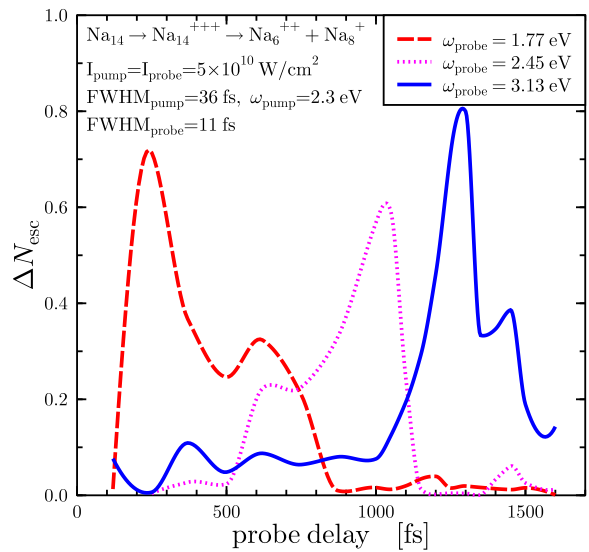,width=7.8cm}
\caption{\label{fig:delay}
Extra ionization due to the probe pulse (as compared to mere pump ionization) 
as a function of delay time for three typical frequencies. One 
clearly spots the three regimes associated to the three frequencies :
low frequency dominant at short delay (full line), medium frequency
dominant for intermediate delays (dotted line) and high frequency
characteristic of post-fission and thus large delays (dashed
line). See also figure~\ref{fig:scheme} for completeness.
}
\end{figure}

First let us start with a standard P\&P analysis by fixing the probe
frequency $\omega_{\rm probe}$ and by varying its time delay.
For the present test case, the Mie frequencies of the two fission fragments
are about 0.7 to 1.5 eV above that of the elongated $\rm
{Na_{14}}^{+++}$ which lies around 1.9 eV. By choosing a probe
frequency at about 3 eV, we should thus be able to track 
down the scission time by scanning the dynamics in terms of 
time delay. Before scission the fragmenting system should appear 
more "red" while afterwards the system will appear more "blue".
Indeed the observable here is the additional
ionization due
to the probe laser with respect to the mere pump ionization, defined
as
$
\Delta N_{\rm esc} = N_{\rm esc}[{\rm probe}] - N_{\rm esc}[{\rm
    pump}].
$
That quantity depends slightly on time for a given delay. We consider
here an 
``asymptotic'' value, recorded 500 fs after the probe pulse.  The
$\Delta N_{\rm esc}$ reflects the closeness of the probe frequency to
that of the plasmon. For delays lower than the scission time,
$\omega_{\rm probe}$ remains far away from the fragment plasmon and
this off-resonance gives vanishing or low extra ionizations. However
as we increase the probe delay steadily, a sudden increase of $\Delta
N_{\rm esc}$ should occur, corresponding to the first resonance of
$\omega_{\rm probe}$ with one or both fragment plasmons. Thus the
first delay at which we observe a significant ionization after probe
laser gives us an estimate of the scission time. This is indeed what
we obtained for $\omega_{\rm probe}=3.13$ eV, see the full curve in
figure~\ref{fig:delay}: the obtained extra ionization, as a function of
delay, is insignificant up to 1 ps and jumps around 1.25 ps.

An alternative strategy is to use a low probe frequency, close to that
of the $\rm {Na_{14}}^{+++}$ compound system, and to check the
disappearance of $\Delta N_{\rm esc}$ with increased delay time.
The picture can be complemented by tracking also an intermediate
frequency.
These two scenarios are shown together with the high frequency
calculations in figure~\ref{fig:delay}. The low frequency of 1.77 eV
(long dashes) attaches to the elongated and still connected
cluster 
while the intermediate frequency (dotted curve) attaches
the final stages of pre-scission.  As expected the three signals are
nicely complementing each other and allow to conclude on a scission
time of order 1.2 ps for the given test case.

\begin{figure}
\epsfig{figure=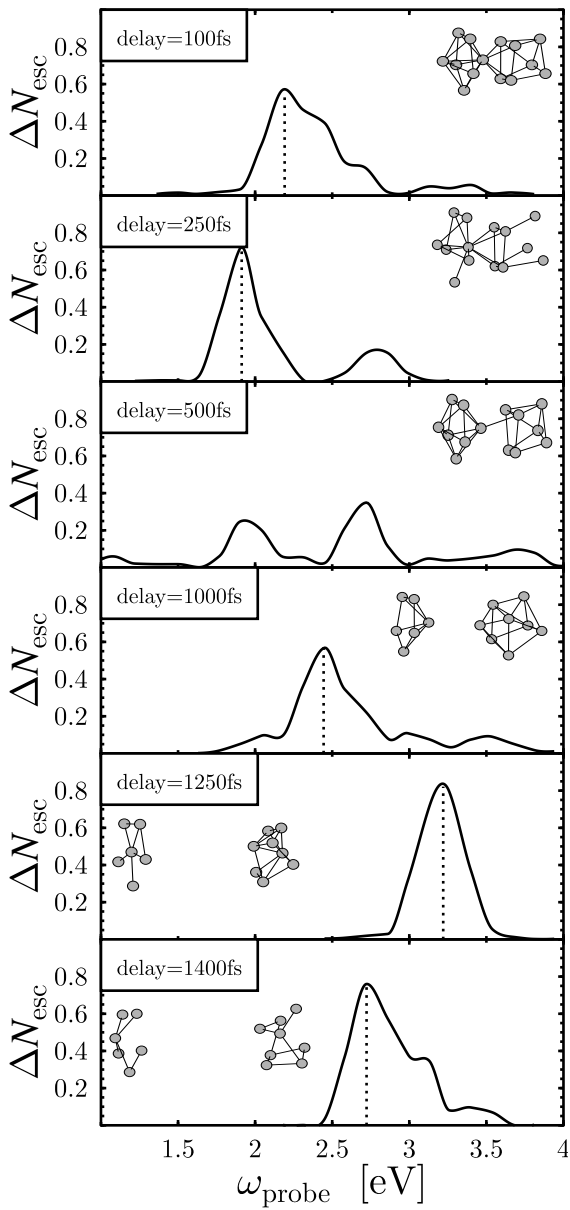,width=8cm}
\caption{\label{fig:freq}
Extra emission due to the probe pulse (as compared to the pump ionization) 
as a function of probe frequency for a few typical delays along 
the fission path. Ionic structures at the corresponding stages during
the process are shown as inserts. 
}
\end{figure}

The full richness of resonance spectra at a given stage of fission can
be studied in even more detail by fixing the time delay and scanning a
dense variety of 
probe laser frequencies. This way of analysis is illustrated in
figure \ref{fig:freq}, which demonstrates that the plasmon peak
position (see vertical lines at the maximum of each spectrum for
guidance) and its fragmentation pattern change dramatically on the way 
to scission.  In turn, a probe pulse with fixed frequency will
come into resonance only during a rather well defined delay window,
as was seen in figure~\ref{fig:delay}. 
A world of interesting detailed information is thus contained in
figure~\ref{fig:freq} when scanning the probe frequency.
Such multi-color P\&P setups are, however, an enormous experimental
challenge.
The trend of the plasmon peak sketched in figure~\ref{fig:scheme}
complies with figure~\ref{fig:freq}. For 500 fs delay, the spectrum is
too fragmented to define unambigously a plasmon peak. This holds for
the transitional interval 300--800 fs, as hinted by a dotted regime in
figure~\ref{fig:scheme}.

\begin{figure}
\epsfig{figure=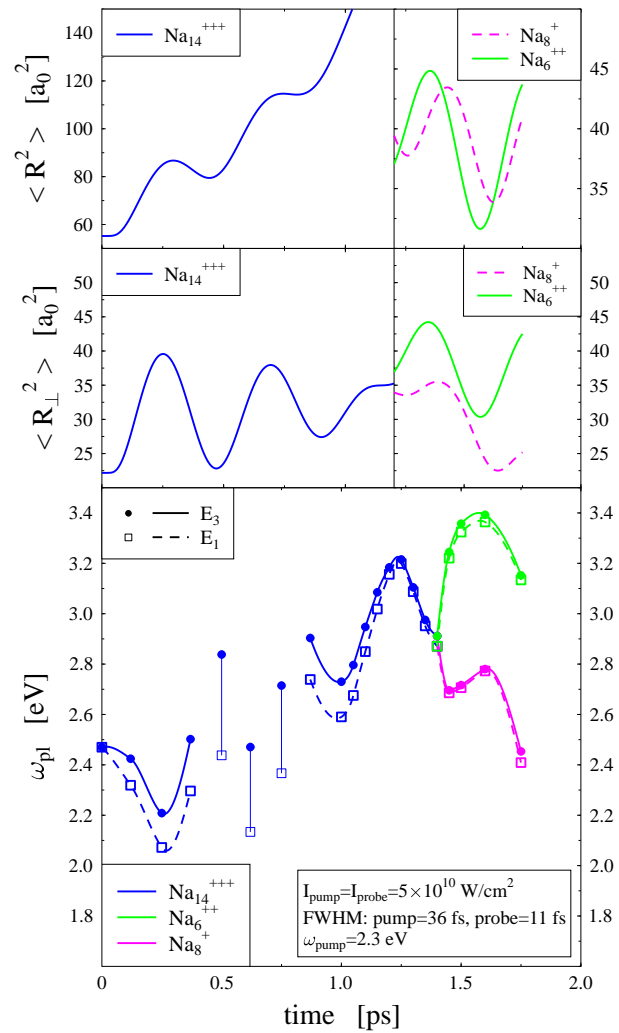,width=8cm}
\caption{\label{fig:plasmon}
Time evolution of various observables related to ionic structure
along the fission path. Bottom: Plasmon frequency estimated from
a moment analysis of the emission distributions (see text for
details). Because of the too spread response, we only indicates values
of $E_1$ and $E_3$ between 300 and 800 fs. Middle: Transverse mean
square radius of the cluster before 1.4 ps and the two fission
fragments for later times. Top: Total ionic mean square radius.}
\end{figure}

Thus tuning probe delay time and frequency complement each other to
analyze the fission process in great detail. A summary analysis can
be performed in terms of moments of the resonance spectra
\cite{Boh79}. To that end, we compute weighted averages over the
ionization spectra 
$ 
E_n = 
\sqrt{{\int\textrm d\omega\,\Delta N_{\rm esc}\,\omega^n}\Big/
{\int \textrm d\omega \, \Delta N_{\rm esc}\,\omega^{n-2}}}.
$
Good reference averages are $E_1$ and $E_3$ which put moderate
emphasis on the lower ($E_1$) and the upper ($E_3$) side of the
spectrum. A well peaked resonance response is signaled by $E_3 \simeq
E_1$. The difference between the two values indicates the width of the
spectrum. The results for $E_1$ and $E_3$ are shown in the bottom
panel of figure~\ref{fig:plasmon}.  As can also be seen  from
figure~\ref{fig:freq}, the spectrum is spread up to 1 ps, which
corresponds to a large difference between $E_1$ and $E_3$. Above that
delay time, both averages coincide. 
At short times, an initial red shift is observed, whereas above 1 ps, a
sudden blue shift occurs up to 1.3 ps, followed by a final red shift:
this panel, compared with figure~\ref{fig:scheme}, shows that the
results are once again consistent with the expected time evolution of
the plasmon. Even if well defined values of $E_1$ and $E_3$ are
  hard to assign in the 300--800 fs delay range, one can see that
the average values oscillate with
a period of a few hundred fs which shrinks somewhat with time going
on. We expect that the mean plasmon frequency depends on the ionic
structure. Thus we check the time evolution of the transverse and the
total mean square radii, respectively $\langle {R_\perp}^2 \rangle$ and
$\langle R^2 \rangle$, see middle and top panels of
figure~\ref{fig:plasmon}. Note that after 1.2 ps, the radii are
calculated 
for the two separate fission fragments relative to each separate
center of mass.  Before scission time, $\langle R^2 \rangle$ shows a
breathing although partially masked by the global extension of the
cluster along its symmetry axis. The transversal radius $\langle
{R_\perp}^2 \rangle$ also oscillates but is not exactly in phase with
$\langle R^2 \rangle$.  Both radii present periods that shrink with
increasing time. We note a relatively good correspondence between the
minima of $\omega_{\rm pl}$ and the maxima of the radii, and {\it vice
versa} for the maxima of $\omega_{\rm pl}$ and the minima of the
radii. The oscillations of the plasmon frequency are not dominated by
either the transverse ionic motion or the longitudinal one, but seem
to be a mixture of both. After scission, the plasmon evolution is
clearly related to the radial oscillations since the extrema of
$\omega_{\rm pl}$ and $\langle {R_\perp}^2 \rangle$ perfectly
coincide. Thus this ``measurement'' of the plasmon frequency through
the resonant ionization of the cluster gives access to the time
evolution of the ionic structure, and to some extent to the
potential energy surface along the fission path.

In this paper we have presented a scenario for pump and probe (P\&P)
analysis of fission of metal clusters. The scheme exploits the marked
response of the Mie surface plasmon to laser fields and the fact that
the plasmon resonance spectrum changes dramatically along the fission
path. The response is quantified in a measurable way in terms of
ionization induced by the laser pulse. With a single-frequency P\&P
setup, one can measure the scission time (that is, the final break up
into fragments) when properly tuning the frequency. Scanning the
frequency of the probe pulses would allow to resolve more details of
the ionic geometries along the fission path.
Finally, we ought to mention that a proper orientation of the cluster
is required during the time span of the measurement time, {\it i.e.}
for a few ps. This can be achieved by 
preparing preliminary a sample of clusters of same orientation, for
instance, with the laser burning technique~\cite{Wen99,Sei00}; and by
cooling the beam to a temperature of about 10 K, so that the
orientation is conserved for at least 2 ps.

Acknowledgments. The authors thank the French-German exchange
program PROCOPE, the Alexander von Humboldt foundation, the French
ministry of education, the CNRS program "Mat\'eriaux" and the Institut
Universitaire de France for financial support during the realization
of this work.

\bibliographystyle{unsrt}

\end{document}